\documentclass[preprint,showpacs,preprintnumbers,amsmath,amssymb]{revtex4}

\usepackage{graphicx}
\usepackage{dcolumn}
\usepackage{bm}
\usepackage{epsfig}

\def\e{\begin{equation}}
\def\f{\end{equation}}
\def\=#1{\overline{\overline #1}}

\def\-#1{{\bf #1}}
\def\.{\cdot}
\def\l#1{\label{eq:#1}}
\def\r#1{(\ref{eq:#1})}

\begin{document}

\title{Superlens made of a metamaterial with extreme effective parameters}

\author{M\'{a}rio G. Silveirinha$^1$}
\altaffiliation[]{Electronic address: mario.silveirinha@co.it.pt}
\author{Carlos A. Fernandes$^2$}
\author{Jorge R. Costa$^{2,3}$}
\affiliation{$^1$University of Coimbra, Department of Electrical
  Engineering-Instituto de Telecomunica\c{c}\~{o}es, 3030 Coimbra,
  Portugal \\
$^2$Technical University of Lisbon, Instituto Superior
T\'{e}cnico-Instituto de Telecomunica\c{c}\~{o}es, 1049-001
Lisbon,Portugal\\
$^3$Instituto Superior de Ci\^{e}ncias do Trabalho e da Empresa,
Departamento de Ci\^{e}ncias e Tecnologias da Informa\c{c}\~{a}o,
1649-026 Lisboa, Portugal}

\date{\today}

\begin{abstract}
We propose a superlens formed by an ultra-dense array of crossed
metallic wires. It is demonstrated that due to the anomalous
interaction between crossed wires, the structured substrate is
characterized by an anomalously high index of refraction and
supports strongly confined guided modes with very short propagation
wavelengths. It is theoretically proven that a planar slab of such
structured material makes a superlens that may compensate for the
attenuation introduced by free-space propagation and restore the
subwavelength details of the source. The bandwidth of the proposed
device can be quite significant since the response of the structured
substrate is non-resonant. The theoretical results are fully
supported by numerical simulations.
\end{abstract}

 \maketitle

\section{Introduction}
The resolution of classical imaging systems is limited by Rayleigh's
diffraction limit, which establishes that the width of the beam spot
at the image plane cannot be smaller than $\lambda/2$. The
subwavelength details of an image are associated with evanescent
spatial harmonics which exhibit an exponential decay in free-space.
Classical imaging systems only operate with propagating spatial
harmonics, and due to this reason the subwavelength information is
irremediably lost.

\begin{figure}[bt] \centering
\epsfig{file=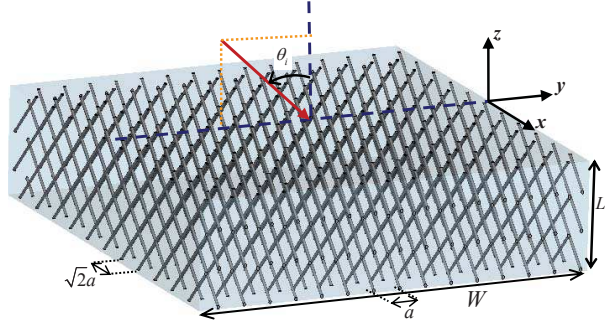, width=8cm} \caption{(Color online)
Geometry of the superlens: an array of crossed metallic wires tilted
by $\pm$45[deg] with respect to the interfaces is embedded in a host
material slab with thickness $L$ and relative permittivity
$\varepsilon_h$. The wires are contained in planes parallel to the
\emph{xoz} plane. Each wire mesh is arranged in a square lattice
with lattice constant $a$. The wires in adjacent planes are
orthogonal, and are spaced by $a/2$.} \label{geomwires}
\end{figure}

In recent years, there has been a great interest in approaches that
may allow overcoming the classical diffraction limit and enable
subwavelength imaging \cite{Pendry_PerfectLens, Taubner, Silversub,
Marques_MagnetoLens, SWIWM, Salandrino, HyperScience, Smolyaninov,
Merlin}. In particular the ``perfect lens'' introduced in
\cite{Pendry_PerfectLens} has received significant attention, due to
its promise of perfect imaging. The superlens is formed by a thin
slab of a material with either negative permittivity \cite{Taubner,
Silversub} or negative permeability \cite{Marques_MagnetoLens}, or
both. The mechanism used by the superlens to restore the evanescent
spatial spectrum is based on the resonant excitation of surface
waves. In this work, we explore an alternative approach to obtain
imaging with subwavelength resolution that does not require
materials with negative effective parameters. In our recent work
\cite{MarioEVL}, it was shown that a structured material formed by
an array of crossed metallic wires (see Fig. \ref{geomwires}) has
very peculiar properties, and is characterized by an anomalously
high positive index of refraction. This property is quite surprising
since intuitively one might expect that in the long wavelength limit
the material would not support propagating plane waves, and instead
would have plasma-like properties. It was demonstrated in
\cite{MarioEVL}, that in reality the material is not characterized
by a negative permittivity, but quite differently it may interact
with electromagnetic waves as a material with very large positive
permittivity. In particular, a slab of the considered material may
support ultra-subwavelength guided modes with very short guided
wavelengths \cite{MarioEVL}. These exciting properties have been
experimentally verified at microwaves in \cite{EVLexp}. Here, we
will demonstrate that the resonant excitation of guided modes may
enable a ``superlensing'' effect somehow analogous to the effect
obtained using a silver lens \cite{Pendry_PerfectLens, Silversub}.

It is important to mention that arrays of parallel metallic wires
have been used before to achieve sub-diffraction imaging
\cite{SWIWM, 1meter}. However, in these previous works the imaging
mechanism was based on the conversion of evanescent waves into
transmission line modes, and did not involve the excitation of
guided modes or the enhancement of evanescent waves. Due to this
reason, in \cite{SWIWM, 1meter} the source must be placed in the
immediate vicinity of the imaging device. Quite differently, in our
current proposal the subwavelength spatial spectrum is restored due
to the excitation of guided modes. This mechanism enables to
compensate for the exponential decay of evanescent waves outside the
lens, and permits that the source and image plane are located at a
distance from the interfaces comparable to the lens thickness. This
is of obvious interest for practical applications, and may enable,
for example, to image ``buried'' objects with subwavelength
resolution.

This paper is organized as follows. In section \ref{SecTransfer},
the geometry of the metamaterial is described and the transfer
function of a metamaterial slab is derived using homogenization
methods. In section \ref{SecEvanescent}, it is demonstrated that
evanescent waves are enhanced by the proposed superlens. In section
\ref{SecLine}, we study the imaging of a line source by the
metamaterial slab. Finally, in section \ref{SecConcl} the conclusion
is drawn. In this work, it is assumed that the electromagnetic
fields are harmonic and have time variation of the form $e^{j \omega
t}$.

\section{Homogenization of the material \label{SecTransfer}}

The geometry of the proposed superlens is depicted in Fig.
\ref{geomwires}. It consists of an array of crossed metallic wires
embedded in a dielectric substrate with relative permittivity
$\varepsilon_h$ and thickness $L$. The metallic wires are parallel
to the $xoz$ plane and have radius $r_w$.

As described in our previous work \cite{MarioEVL}, the considered
structure can be conveniently analyzed using homogenization
techniques. To begin with, we consider a simple plane wave
scattering problem such that the incoming electric field is
polarized along the $x$-direction, and the incident wave vector is
in the $yoz$ plane, as illustrated in the inset of Fig. \ref{kscan}.
Similar to the results of \cite{MarioEVL}, the electric field in all
space only has an $x$-component and can be written as (the
$y$-dependence of the fields is suppressed):
\begin{eqnarray}
E_x  &=& E_x^{inc} \left( {e^{\gamma _0 z}  + R e^{ - \gamma _0
z}} \right),\quad \quad z > 0 \nonumber \\
E_x &=&  {A^{+}_1  e^{ -j {\kern 1pt} k_z^{\left( 1 \right)} z}  +
A^{-}_1  e^{ +j{\kern 1pt} k_z^{\left( 1 \right)} z} +
} \nonumber \\
&& {A^{+}_2  e^{ -j {\kern 1pt} k_z^{\left( 2 \right)} z}  + A^{-}_2
e^{ +j{\kern 1pt} k_z^{\left( 2 \right)} z} },\quad  -L< z < 0 \nonumber \\
E_x  &=& E_x^{inc} T e^{+\gamma _0 z},\quad \quad z < -L
\end{eqnarray}
where $E_x^{inc}$ is the incident field, $\gamma _0  = \sqrt {k_y^2
- \omega ^2 \varepsilon _0 \mu _0 }$, $k_y = \omega \sqrt
{\varepsilon _0 \mu _0 }\sin \theta_i $ only depends on the angle of
incidence $\theta_i$, and $R$ and $T$ are the reflection and
transmission coefficients, respectively. The propagation constants
$k_z^{\left( 1,2 \right)}$ and the amplitudes $A^{\pm}_i$ are
associated with the modes excited inside the metamaterial slab. As
discussed in Refs. \cite{Silv_AP_WM, Silv_MTT_3DWires,
Constantin_WM, MarioEVL}, the considered material is strongly
spatially dispersive and consequently it supports two
electromagnetic modes with the same electric field polarization, but
with different propagation constants. The propagation constants
$k_z^{\left( 1,2 \right)}$ of these modes are obtained from the
solution with respect to $k_z$ of the dispersion equation
\cite{MarioEVL}: \e \left(1 + \frac{1}{{\frac{\varepsilon_h}{{\left(
{\varepsilon_m - \varepsilon_h} \right)f_V }} - \frac{{\beta^2 -
k_z^2/2 }}{{\beta _p^2 }}}} \right) \beta^2 = k_y^2 + k_z^2
\l{dispeq}\f where $ \beta _p = \left[ {2\pi /\left( {\ln \left(
{a/2\pi r_w } \right) + 0.5275} \right)} \right]^{1/2} /a$ is the
plasma wavenumber, $f_V = \pi (r_w/a)^2$, $\beta = \omega \sqrt
{\varepsilon _h \varepsilon_0 \mu _0 }$ is the wavenumber in the
host medium, and $\varepsilon_m$ is the complex relative
permittivity of the metal. The dispersion equation is equivalent to
a quadratic equation in the variable $k_z^2$. In the particular case
where the metallic wires are perfect electric conductors (PEC), i.e.
$\varepsilon_m = - \infty$, the solutions are:
\begin{eqnarray}
k_z^2 = \frac{1}{2}\left( {3 \beta^2 - k_y^2 \pm \sqrt
{\beta^4 + \beta^2 \left( {2k_y^2  + 8\beta _p^2 } \right) + k_y^4 } } \right), \nonumber \\
\rm{(PEC\,\,wires)}\,\, \l{kzWM}
\end{eqnarray}
In order to calculate the reflection and transmission coefficients,
it is necessary to impose the following boundary conditions:
\begin{eqnarray}
E_x \,\,{\rm{and}}\,\,\frac{{dE_x
}}{{dz}}\,\,{\rm{are\,\,continuous\,\,at}}\,\,z =
0\,{\rm{ and }}\,z =-L \nonumber \\
\frac{{d^2 E_x }}{{dz^2 }} + \left( {\beta^2 - k_y^2 } \right)E_x =
0, \,\,\, z = 0^{-}\,{\rm{ and }}\,z = -L^{+}
 \l{BCs}
\end{eqnarray}
\begin{widetext}
The first set of boundary conditions corresponds to the continuity
of the tangential electromagnetic fields $E_x$ and $H_y$ at the
interfaces, while the second set of boundary conditions is necessary
to properly model the effects of spatial dispersion and the
microscopic behavior of the electric currents along the wires
\cite{MarioEVL,ABCtilted}. Using the proposed homogenization model
it can be proved that the transmission coefficient is given by:
\begin{eqnarray}
 T\left( {\omega ,k_y } \right) = \frac{1}{{1 - \frac{{k_z^{\left( 1 \right)} }}{{\gamma _0 }}\frac{{\left( {k_z^{\left( 2 \right)} } \right)^2  + \gamma _h^2 }}{{\left( {k_z^{\left( 2 \right)} } \right)^2  - \left( {k_z^{\left( 1 \right)} } \right)^2 }}\tan \left( {\frac{{k_z^{\left( 1 \right)} L}}{2}} \right) + {\kern 1pt} \frac{{k_z^{\left( 2 \right)} }}{{\gamma _0 }}\frac{{\left( {k_z^{\left( 1 \right)} } \right)^2  + \gamma _h^2 }}{{\left( {k_z^{\left( 2 \right)} } \right)^2  - \left( {k_z^{\left( 1 \right)} } \right)^2 }}\tan \left( {\frac{{k_z^{\left( 2 \right)} L}}{2}} \right)}} \nonumber \\
 {\rm{              }} - \frac{1}{{1 + \frac{{k_z^{\left( 1 \right)} }}{{\gamma _0 }}\frac{{\left( {k_z^{\left( 2 \right)} } \right)^2  + \gamma _h^2 }}{{\left( {k_z^{\left( 2 \right)} } \right)^2  - \left( {k_z^{\left( 1 \right)} } \right)^2 }}\cot \left( {\frac{{k_z^{\left( 1 \right)} L}}{2}} \right) - \frac{{k_z^{\left( 2 \right)} }}{{\gamma _0 }}\frac{{\left( {k_z^{\left( 1 \right)} } \right)^2  + \gamma _h^2 }}{{\left( {k_z^{\left( 2 \right)} } \right)^2  - \left( {k_z^{\left( 1 \right)} } \right)^2 }}\cot \left( {\frac{{k_z^{\left( 2 \right)} L}}{2}} \right)}}
\l{Tcoeff}
\end{eqnarray}
\end{widetext}
where $\gamma _h  = \sqrt {k_y^2 - \omega ^2 \varepsilon_h
\varepsilon _0 \mu _0 }$. As it will be made clear ahead, $ T\left(
{\omega ,k_y } \right)$ may be regarded as the ``optical'' transfer
function of the superlens.

\begin{figure}[thb] \centering
\epsfig{file=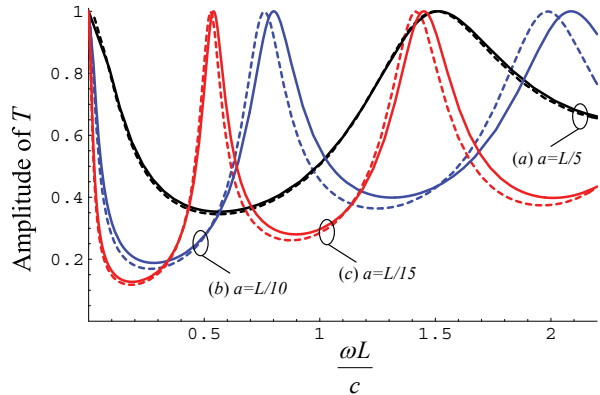, width=8cm} \caption{(Color online) Amplitude
of the transmission coefficient as a function of normalized
frequency for different lattice constants, $a$, and a fixed slab
thickness $L$. Solid lines: analytical model; Dashed lines: full
wave simulations obtained with CST Microwave Studio \cite{CST2006}.
The angle of incidence is $\theta_i = 15^o$, the radius of the wires
is $r_w=0.05a$, and the permittivity of the substrate is
$\varepsilon_h=1$.} \label{Tfreq}
\end{figure}
To begin with, we study the frequency response of the metamaterial
slab for plane wave incidence. In Fig. \ref{Tfreq} the transmission
coefficient $T$ is plotted as function of frequency for a fixed
thickness $L$ and for different values of the lattice constant $a$:
$a=L/5$, $a=L/10$ and $a=L/15$ (solid lines). The wires are assumed
to be PEC, and the angle of incidence is $\theta_i = 15^o$. In all
the examples, the amplitude of the transmission coefficient is
approximately unity in the static limit $\omega = 0$. This is
consistent with the fact that for low frequencies the length of the
wires ($L_w = \sqrt 2 L$) is electrically short, and thus at $\omega
= 0$ the wires may interact weakly with the incoming wave. However,
quite remarkably and despite the fact that the wires are very short,
for frequencies slightly above $\omega = 0$ the transmission
coefficient may have several dips and the incoming wave may be
strongly reflected by the structured substrate. For example, when
the lattice constant is $a=L/15$ (curve (c)), $T$ has a dip for
$\omega L/c \approx  0.2$, which corresponds to metallic wires with
length $L_w = 0.04 \lambda_0$. This length is one order of magnitude
lower than the traditional $0.5 \lambda_0$ resonance length for a
PEC wire. The unusual electrical response of the crossed wire mesh
(formed by very short wires) is caused by the strong interaction
between the fields guided by each set of wires and the other set of
perpendicular wires, which results in an anomalously high positive
effective index of refraction \cite{MarioEVL}.

\begin{figure}[t] \centering
\epsfig{file=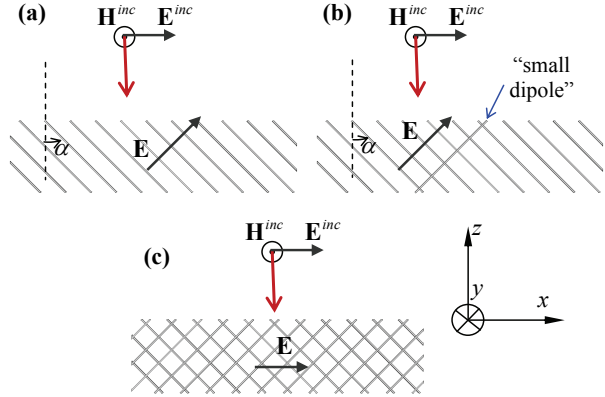, width=8cm} \caption{(Color online) Step by
step construction of the crossed wire mesh [panel (c)] starting from
an array of parallel wires tilted by an angle $\alpha$ with respect
to the interface [panel (a)]. The incoming wave propagates along the
normal direction and the wires stand in air. In order to delay the
wave supported by the structure of panel (a) one can add suitable
inclusions to the material (``small wire dipoles'') [panel (b)].
These inclusions will partially block the path of propagation, and
in this way increase the effective index of refraction of the
composite material. By adding many of such ``wire dipoles'' we
arrive at the configuration of panel (c).} \label{physexp}
\end{figure}
This mechanism can be understood by analyzing Fig. \ref{physexp},
where the crossed wire mesh [panel (c)] is constructed step by step
starting from an array of parallel wires [panel (a)], to which the
perpendicular wires are sequentially added [panel (b)]. For
simplicity, it is supposed that in the scenarios depicted in Fig.
\ref{physexp} the incoming wave illuminates the slab along the
normal direction, and that the host material is air. Consider first
the configuration reported in panel (a), which portrays the case
where the array of parallel wires is tilted by an angle $\alpha$
(not necessarily $45^o$) with respect to the interface. It is known
from previous studies \cite{ABCtilted} that when the wires are
densely packed ($a/L_w \ll 1$) the dominant electromagnetic mode
excited in such structure is the so-called ``transmission line''
mode \cite{WMPRB}. This mode is a transverse electromagnetic (TEM)
wave, and hence the average electric field inside artificial
material is orthogonal to the wires, as represented in panel (a).
The fields inside the material propagate along the direction of the
wires with the velocity of light, and the propagation constant along
the $z$ direction is $k_z = \sec \alpha \, \omega/c $ (assuming that
the incoming wave propagates along the normal direction)
\cite{ABCtilted}. Hence, $k_z c/\omega \sim 1$, at least as long as
$\alpha$ is not too large, which is the situation of interest here.
How can we increase $k_z c/\omega $, so that the effective index of
refraction seen by the wave in the material may be very large? A
possible solution is to add suitable inclusions to the metamaterial
so that the path of the transmission line wave is partially
obstructed by these inclusions resulting in a greater phase delay.
Since from panel (a) it is known that the electric field is
orthogonal to the wires, the obvious solution is to add a ``small
wire dipole'' that will partially block the electric field
associated with the original transmission line mode [solution
represented in panel (b)]. Naturally, in order to obtain a
significant phase delay many of such ``small wire dipoles'' are
necessary, and hence we arrive at the completely symmetric crossed
wire mesh configuration shown in panel (c). It is important to note
that the array of ``added wires'' does also support transmission
line modes when it stands alone in free-space. However, quite
interestingly, such TEM wave is delayed by the original set of wires
[represented in panel (a)], precisely in the same manner as the
``added wires'' delay the TEM wave supported by the original set of
wires. In fact, when $\alpha = 45^o$ the crossed wire mesh becomes
completely symmetric and the roles of two sets of wires can be
interchanged. It is also clear that for symmetry reasons the average
electric field inside the artificial material becomes parallel to
the incident field when the second array of wires is added to the
structure. The described interaction mechanism is the physical
reason for the very large index of refraction of the crossed wire
mesh. It is important to mention that the crossed wire mesh supports
a propagating mode at very low frequencies (with the electric field
polarized along the $x$-direction) only when the two sets of wires
are physically disconnected, as assumed here. Indeed, when the two
sets of wires are connected it is possible to find a continuous
(``zigzag'') path along the metallic wires that runs from
$x=-\infty$ to $x=+\infty$. This continuous metallic channel would
effectively block an incoming wave (with the electric field along
the $x$-direction), and confer the material plasmonic-like
properties. Quite differently, when the two sets of wires are
nonconnected, any continuous path running along the wires has finite
length, and consequently propagation becomes possible at low
frequencies.

As shown in \cite{MarioEVL}, for a fixed frequency the effective
index of refraction of the propagating mode, $n_{ef} = k_z
c/\omega$, ($k_z$ is given by the solution of Eq. \r{kzWM}
associated with the ``+'' sign), can be made arbitrarily large by
increasing the density of wires, i.e. by making $a \to 0$. Such
phenomenon can be readily understood in light of the interaction
mechanism explained in the previous paragraph. In fact, as the
density of wires increases the wave guided by each set of wires
suffers a greater phase delay, resulting in a larger effective index
of refraction. The results of Fig. \ref{Tfreq} are consistent with
this property, since as $a/L$ is made smaller the electric response
of the structure becomes stronger for lower frequencies. In fact,
notwithstanding the extremely small electrical length of the wires,
theoretically it is possible to move the first dip of transmission
coefficient to an arbitrarily small frequency by reducing $a$ and
keeping the metal volume fraction constant ($r_w/a = const.$). This
property is only limited by the effect of metallic loss, which may
be relevant when the radius of the wires becomes smaller than the
skin depth of the metal \cite{MarioEVL}. The reported results are
fully confirmed by full wave simulations obtained using the
electromagnetic simulator CST Microwave Studio$^{\rm{TM}}$
\cite{CST2006} (dashed lines in Fig. \ref{Tfreq}), which show a
remarkable agreement with our analytical model.

\section{Enhancement of evanescent waves \label{SecEvanescent}}

\begin{figure}[thb] \centering
\epsfig{file=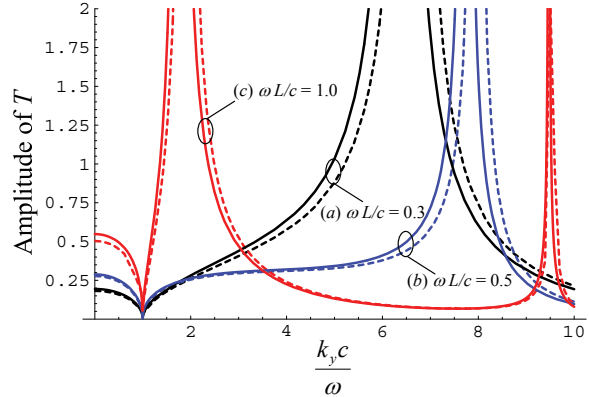, width=8cm} \caption{(Color online)
Amplitude of the transmission coefficient as a function of the
normalized transverse wave vector, $k_y$, for different frequencies
of operation $\omega$, and a fixed thickness $L$ of the material
slab. Notice that for $k_y c/\omega
>1$ the incoming wave is an evanescent mode, whereas if $\left|k_y
c/\omega\right| <1$ the incoming wave is a propagating plane wave.
The lattice constant is $a=L/10$, the radius of the PEC wires is
$r_w = 0.05a$, and $\varepsilon_h=1.0$. Solid lines: analytical
model; Dashed lines: full wave simulations obtained with CST
Microwave Studio \cite{CST2006}.} \label{kscanPEC}
\end{figure}

The high index of refraction of the material may enable the
propagation of ultra-subwavelength guided modes \cite{MarioEVL,
EVLexp}. This suggests that a thin material slab may be used to
enhance evanescent waves, and in this way obtain a superlensing
effect. In order to investigate this possibility in Fig.
\ref{kscanPEC} we plot the amplitude of $T$ as a function of the
transverse wavenumber of the impinging wave, $k_y$, for different
frequencies of operation $\omega$. The wires are assumed to be PEC
and the slab has a fixed thickness $L$. The solid curves correspond
to the analytical model [Eq. \r{Tcoeff}], whereas the dashed lines
were calculated using CST Microwave Studio$^{\rm{TM}}$
\cite{CST2006}.
It is seen that the transfer function $T=T(k_y)$ may have several
resonances, which correspond to the excitation of guided modes. For
very low frequencies (not shown in Fig. \ref{kscanPEC}), the pole of
$T=T(k_y)$ occurs for some transverse wavenumber $k_y$ such that
$k_y c/\omega$ is slightly above unity, consistently with the fact
that near the static limit most of the energy of the guided mode is
concentrated outside the material slab. The remarkable property is
that due to extreme effective parameters of the composite material
the guided mode may become extremely attached to the metamaterial
slab even for extremely long wavelengths. For example, in curve (a)
the pole occurs at $k_y c/\omega \approx 6.3$, notwithstanding the
extremely small thickness of the slab thickness, $L=0.048
\lambda_0$, and the very small length of the wire inclusions ($L_w =
0.068 \lambda_0$). As illustrated in curve (b), as the frequency
increases the value of $k_y$ associated with the guided mode
increases even more, which indicates that the mode becomes even more
attached to the structured substrate. For even larger frequencies
[curve (c)], an additional guided mode is supported, and as a
consequence two distinct resonances are seen in Fig. \ref{kscanPEC}.

From the results of Fig. \ref{kscanPEC}, it may be inferred that the
optimal frequency to obtain subwavelength imaging is such that
$\omega L/c \approx 0.3$ [curve (a)]. In fact, for such frequency
the evanescent waves may be significantly amplified over a
relatively wide range of values of $k_y$ centered at the pole of
$T(k_y)$. It should be noticed that at the same frequency the
amplitude of the transmission coefficient for propagating waves
($k_y c/\omega<1$) is relatively small. This further helps to
enhance the subwavelength spatial spectrum (which is attenuated in
the free-space regions) in comparison to propagating plane waves
(which propagate with no attenuation in free-space). Indeed, from
curve (b) of Fig. \ref{Tfreq} it may be verified that $\omega L/c
\approx 0.3$ corresponds to a minimum of $T$ for incidence along the
broadside direction. The results of our simulations indicate that
the described property is also observed for other lens parameters,
and that the value of $\omega L/c$ that enables the best imaging is
such that $T$ has a minimum for normal incidence.
\begin{figure}[th] \centering
\epsfig{file=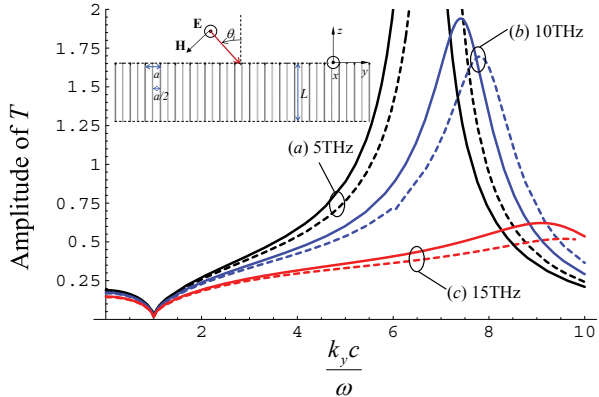, width=8cm} \caption{(Color online)
Amplitude of the transmission coefficient as a function of the
normalized transverse wave vector, $k_y$, for the frequencies of
operation 5THz, 10THz, and 15THz, and for a material slab with
thickness $L=2.9\mu m$, $L=1.4\mu m$, and $L=0.95\mu m$,
respectively (thus $L = 0.3 c/\omega$ in all the cases). The lattice
constant is $a=L/10$, the radius of the silver wires is $r_w =
0.05a$, and $\varepsilon_h=1.0$. Solid lines: analytical model;
Dashed lines: full wave simulations \cite{CST2006}. The inset
represents the geometry of the artificial material slab.}
\label{kscan}
\end{figure}

The enhancement of evanescent waves has a good tolerance to the
effect of metallic loss. In order to demonstrate this, we plot in
Fig. \ref{kscan} the transfer function of a system with the same
parameters as the one associated with curve (a) of Fig.
\ref{kscanPEC}, except that now the effect of metallic loss if fully
considered. The wires are assumed to be made of silver, which is
modeled using a Drude dispersion relation, consistent with the
experimental results tabulated in \cite{OptConst1}. It can be
noticed that at 5THz [curve (a) in Fig. \ref{kscan}] the transfer
function of the system formed by silver wires is rather similar to
that of a system formed by PEC wires [curve (a) in Fig.
\ref{kscanPEC}]. Indeed, at 5THz the skin depth of silver is
approximately $\delta_{Ag} \approx 27nm$ while the radius of the
wires is $r_w = 14nm$, and since $\delta_{Ag}/r_w$ is relatively
close to unity the effect of metallic loss is still quite moderate.
As mentioned before, the effect of metallic loss is negligible
provided $r_w > \delta_{Ag}$ \cite{MarioEVL}. At 10THz [curve (b)]
the response of the system becomes distinctively different from the
PEC case, consistent with the fact that the ratio $\delta_{Ag}/r_w$
increases [$\delta_{Ag} \approx 23nm$ and $r_w = 7nm$]. At 15THz
[curve (c)] the effect of loss is strong [$\delta_{Ag} \approx 23nm$
and $r_w = 5nm$] and there is no amplification of evanescent modes.
Thus, we may conclude that the proposed superlens may be feasible at
least up to 10THz. For frequencies below 5THz the conducting
properties of silver improve and the transfer function of a scaled
superlens formed by silver wires becomes nearly coincident with that
of a superlens formed by PEC wires.

\section{Imaging a line source \label{SecLine}}

\begin{figure}[th] \centering
\epsfig{file=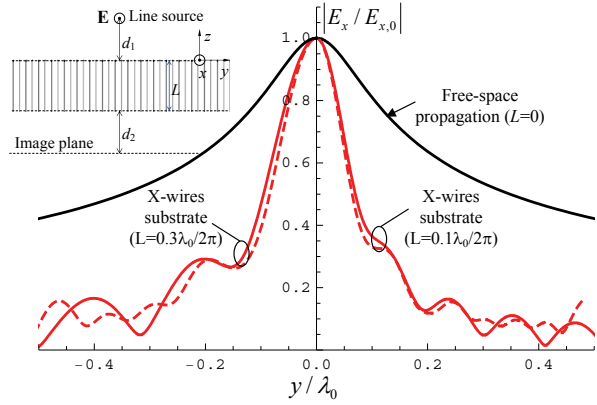, width=8cm} \caption{(Color online)
Amplitude of the normalized electric field imaged by the superlens
for a material slab with: (i) (left-hand side panel, $y<0$) $L=0.3
\lambda_0/2\pi$, $a=L/10$, $r_w=0.05a$, $\varepsilon_h =1$. (ii)
(right-hand side panel, $y>0$) $L=0.1 \lambda_0/2\pi$, $a=L/20$,
$r_w=0.05a$, $\varepsilon_h =1$. Solid black line: electric field
profile in the absence of the metamaterial slab. Solid red (ligth
gray in grayscale) line: image profile for an infinitely extended
metamaterial slab; this curve was calculated using the analytical
model [Eq. \r{Ex_line}]. Dashed red (ligth gray in grayscale) line:
image profile calculated using a full wave numerical method (MoM)
that takes into account all the fine details of the microstructure
of the artificial material for a substrate with finite width $W$
along $y$, given by $W=2.15 \lambda_0$ and $W=0.96 \lambda_0$, for
cases (i) and (ii), respectively. The inset represents the geometry
of the problem: an electric line source is placed a distance $d_1$
above the superlens, and the observation plane is located at a
distance $d_2$ below the lens. For both examples, $d_1=d_2=
0.04\lambda_0$. The relative electric field amplitude with respect
to the case in which the superlens is removed (black curve) is $(a)$
infinite substrate: 0.25 (case i) and 0.17 (case ii). $(b)$ finite
substrate: 0.28 (case i) and 0.19 (case ii).} \label{lineimaging}
\end{figure}

To further characterize the imaging properties of the considered
structure, in the following we study the canonical problem where an
electric line source (infinitely extended along the $x$-direction)
is placed at a distance $d_1$ above the crossed wires slab (see the
inset of Fig. \ref{lineimaging}). The electric field radiated by the
line source is of the form $ E_x = E_0 \frac{1}{{4j}}H_0^{\left( 2
\right)} \left( {\frac{\omega }{c}\rho } \right)$, where $E_0$ is
some constant that depends on the line current, $\rho$ is the radial
distance to the source, and $ H_0^{\left( 2 \right)} = J_0 - j Y_0$
is the Hankel function of second kind and order zero. Assuming that
the metamaterial slab is unbounded along the directions $x$ and $y$,
it is straightforward to find that the electric field at a distance
$d_2$ below the slab is given by the Sommerfeld-type integral: \e
E_x \left( y \right) = \frac{{E_0 }}{\pi }\int\limits_0^\infty
{\frac{1}{{2\gamma _0 }}e^{ - \gamma _0 \left( {d_1  + d_2 }
\right)} \cos \left( {k_y y} \right)T\left( {\omega ,k_y }
\right)dk_y } \l{Ex_line} \f where $\gamma _0  = \sqrt {k_y^2 -
\omega ^2 \varepsilon _0 \mu _0 }$, and $T$ is the transfer function
of the slab given by Eq. \r{Tcoeff}.
Using the above formula we have calculated the electric field
profile at the image plane for an artificial material slab with the
same parameters as curve (a) in Fig. \ref{kscanPEC} ($L = 0.3
\lambda_0/2 \pi$ and $a=L/10$). It was assumed that
$d_1=d_2=0.8L=0.04\lambda_0$, and that the source is in the plane
$y=0$. The normalized electric field at the image plane is depicted
in the left-hand side panel of Fig. \ref{lineimaging} (solid red
line) as a function of $y/\lambda_0$. The half power beam width
(HPBW) is equal to HPBW=$0.12\lambda_0$, which is 4 times smaller
than the traditional diffraction limited value. The normalized
electric field at the image plane when the superlens is absent and
the distance between the source and the image plane is $d_1+d_2$, is
also shown in Fig. \ref{lineimaging} (solid black line). Notice that
when the lens is present the total distance between the source and
the image plane is $d_1+d_2+L=0.124 \lambda_0$, while when the lens
is absent the distance is reduced to $d_1+d_2 = 0.077 \lambda_0$.
Notwithstanding the greater proximity between the source and the
image plane, in the latter case the half power beam width increases
to HPBW=$0.30\lambda_0$. For the propagating distance $d_1+d_2+L$ in
free-space, the half power beam width is HPBW=$0.46\lambda_0$. These
properties demonstrate that the considered artificial material slab
behaves as a superlens with effective ``negative length''. In fact,
the insertion of the superlens in between the source and the image
plane results in a significantly narrower beam than when the lens is
absent, as if the total distance between the source and the image
plane had been decreased. This effect stems from the resonant
excitation of guided modes, which effectively compensate for the
attenuation of evanescent waves in the free-space regions, and in
this way enable imaging with subwavelength resolution. The
resolution of the proposed superlens is $\lambda_0/8$, despite the
source and observation planes being located at a significant
distance from the lens interfaces.

A slab with a larger density of wires may provide an even better
resolution. This is illustrated in the right-hand side panel of Fig.
\ref{lineimaging} (solid red line), for the case in which $a=L/20$,
and $L = 0.1 \lambda_0/2 \pi$. The distances between the source and
image planes are kept the same as in the previous example, so that
$d_1=d_2=2.4L=0.04\lambda_0$. The increased density of wires results
in a larger effective index of refraction of the slab, and this
enhances the response of the system ($T$) for a wider range of
evanescent waves. As a consequence, the half power beam width of the
superlens is improved to HPBW=$0.09 \lambda_0$. The value of HPBW
may obviously be made much smaller if $(d_1+d_2)/L$ is taken closer
to unity, as in a standard silver lens \cite{Pendry_PerfectLens,
Silversub}.

The described results were obtained using the analytical model [Eq.
\r{Ex_line}], and assume that the structured material slab is
infinitely extended along the $x$ and $y$ directions. In order to
confirm the homogenization results, a finite width superlens was
modeled using a dedicated periodic Method of Moments (MoM) code. The
Method of Moments fully takes into account all the fine details of
the microstructure of the crossed wire mesh. In the MoM simulation
the artificial material slab was assumed periodic along the
$x$-direction, and finite along the $y$-direction. The width of
superlens was taken equal to $W=2.15 \lambda_0$ (901 rows of wires
along the $y$ direction spaced by $a/2$) in the case where the
thickness of the lens is $L = 0.3 \lambda_0/2 \pi$, and $W=0.96
\lambda_0$ (2401 rows of wires along the $y$ direction) in the case
where $L = 0.1 \lambda_0/2 \pi$. The calculated electric field
profiles at the image plane are depicted in Fig. \ref{lineimaging}
(dashed red line) showing a remarkable agreement with the
homogenization results, despite that the homogenization results
refer to an unbounded substrate, whereas the MoM simulations refer
to finite width substrate. The results of our simulations (not shown
here for brevity) suggest that this property may also hold for other
widths of the substrate such that $W\approx n \lambda_0$, where $n$
is an integer. On the other hand, we found out that when $W$ is
slightly larger than an odd multiple of $\lambda_0/2$ the reflection
of surface waves at the edges of the slab may significantly corrupt
the quality of the imaging.

\begin{figure}[th] \centering
\epsfig{file=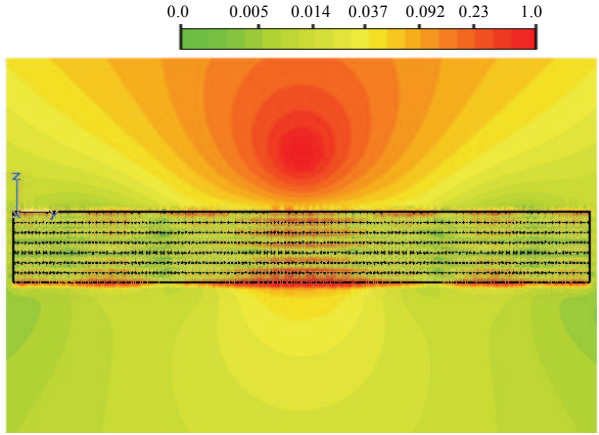, width=8cm} \caption{(Color online)
Amplitude of the normalized electric field $E_x$ for a superlens
configuration with the same geometry as case (i) of Fig.
\ref{lineimaging}, except that the width of the lens along the
$y$-direction is $W=0.39 \lambda_0$.} \label{imagingCST}
\end{figure}
We have also studied the performance of the proposed superlens using
the commercial electromagnetic simulator CST Microwave
Studio$^{\rm{TM}}$ \cite{CST2006}. The geometry of the structure was
taken the same as that of case (i) in Fig. \ref{lineimaging}, except
that (due to limited computational resources) the width of slab
along $y$ was reduced to $W=0.39 \lambda_0$ (162 rows of wires). The
amplitude of the electric field is represented in Fig.
\ref{imagingCST}, showing that despite the relatively small aperture
of the lens the beam width (measured along the along the
$y$-direction) of the transmitted wave is very subwavelength,
consistent with the MoM simulations reported in Fig.
\ref{lineimaging}. The imaging dynamics can be visualized in the
electric field animation reported in \cite{EPAPS}. It is interesting
to note that despite the propagating spatial harmonics not being
matched to the lens, the effects of reflection are hardly
perceptible in the field animation. This happens because the
distance between the source and the lens is electrically small ($d_1
=0.04 \lambda_0$), and interference effects are difficult to
visualize within such short distance.

\begin{figure}[thb] \centering
\epsfig{file=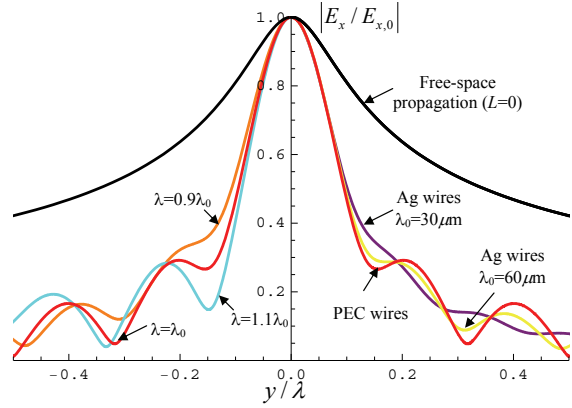, width=8cm} \caption{(Color online)
Amplitude of the normalized electric field imaged by a superlens
characterized by $L=0.3 \lambda_0/2\pi$, $a=L/10$, $r_w=0.05a$,
$\varepsilon_h =1$ at the reference wavelength $\lambda_0$. The line
source is placed at a distance $d_1 = 0.8L$ above the superlens, and
the observation plane is at a distance $d_2 = 0.8L$ below the
superlens. Solid black line: electric field profile in the absence
of the metamaterial slab at $\lambda = \lambda_0$. Solid red (ligth
gray) line: electric field profile for a superlens made of PEC wires
at $\lambda = \lambda_0$. (i) (left-hand side panel, $y<0$) electric
field profile for PEC wires at $\lambda = 0.9 \lambda_0$ and
$\lambda = 1.1 \lambda_0$. (ii) (right-hand side panel, $y>0$)
electric field profile for silver wires at $\lambda_0 = 30 \mu m$
and $\lambda_0 = 60 \mu m$.} \label{imagFreqLoss}
\end{figure}

The described superlensing effect is relatively insensitive with
respect to variations of the frequency of operation. This is shown
in the left-hand side panel of Fig. \ref{imagFreqLoss}, where it is
seen that for the same design as in case (i) of Fig.
\ref{lineimaging} (for an unbounded substrate) the electric field
profile remains nearly invariant when the wavelength of operation is
taken either $\lambda = 0.9 \lambda_0$ or $\lambda = 1.1 \lambda_0$.
This is a consequence of the non-resonant nature of the effective
parameters of the structured material \cite{MarioEVL}. The variation
of the peak level of $\left|E_x\right|$ with respect to the case
$\lambda = \lambda_0$ is less than 8 percent in both cases. The
imaging quality is significantly deteriorated only when the
frequency of operation is such that $\omega L/c$ corresponds to a
maximum of $T$ for plane wave incidence along the broadside
direction.

Likewise, the imaging properties have a good tolerance to the effect
of metallic loss. In the right-hand side panel of Fig.
\ref{imagFreqLoss} the profile of the imaged electric field is shown
for a scaled superlens made of silver wires at $\lambda_0 = 60 \mu
m$ (5THz) and $\lambda_0 = 30 \mu m$ (10THz). Consistent with the
results of Fig. \ref{kscan}, it is seen that apart from a reduced
ripple in the electric field tail (due to the greater attenuation of
the guided modes supported by the metamaterial slab), the half power
beamwidth remains very similar to the PEC case. The relative
amplitude of the transmitted wave with respect to the PEC case is
0.96 at 5THz and 0.84 at 10THz. This suggests that the proposed
structure may have exciting applications at terahertz frequencies.

\section{Conclusion \label{SecConcl}}

It was demonstrated that a crossed wire mesh slab may enable a
imaging with subwavelength resolution. It was discussed that the
extreme effective parameters of the crossed wire mesh permit the
propagation of ultra-subwavelength guided modes. These guided modes
may be excited by the near-field evanescent spectrum of a given
source, and this mechanism, somehow analogous to that characteristic
of a silver superlens \cite{Pendry_PerfectLens, Silversub}, may
permit the enhancement of evanescent waves and of the near field
details. We presented an analytical model that describes very
accurately the electromagnetic response of the proposed imaging
device. For a specific design, and despite the distance between the
source and observation planes and the lens interfaces being
comparable to the lens thickness, the resolution $\lambda_0/8$ was
demonstrated. The effect of metallic loss was analyzed, and it was
shown that it may be relatively mild provided the radius of the
metallic rods is kept smaller than the metal skin-depth. It is
envisioned that the proposed system may find interesting
applications at the microwave and terahertz regimes.

\section{Acknowledgment}
This work is supported in part by Funda\c c\~ao para a Ci\^encia e a
Tecnologia, grant number No. POSC/EEACPS/61887/2004.

\bibliography{imagingEVL}

\end{document}